\documentstyle[12pt]{article}
\begin{document}

\begin{center}
{\small \bf DIRECT T-MATRIX APPROACH TO DETERMINATION OF THE\\ ELECTRIC 
POLARIZABILITIES OF QUANTUM BOUND SYSTEMS} \\[.4in]

\setlength{\baselineskip}{0.1in}
{\footnotesize V. F. KHARCHENKO} \\[.05in]

{\footnotesize \it Bogolyubov Institute for Theoretical Physics,\\
National Academy of Sciences of Ukraine, UA - 03143, Kyiv, Ukraine\\
vkharchenko@bitp.kiev.ua} \\[.4in]
\end{center}

\begin{abstract}
\setlength{\baselineskip}{0.1in}
\noindent
{\footnotesize A new general formalism for determining the electric multipole
polarizabilities of quantum (atomic and nuclear) bound systems based on the 
use of the transition matrix in momentum space has been developed. 
As distinct from the conventional approach with the application of the 
spectral expansion of the total Green's function, our approach does not 
require preliminary determination of the entire unperturbated spectrum; 
instead, it makes possible to calculate the polarizability of a few-body 
bound complex directly based on solving integral equations for the wave 
function of the ground bound state and the transition matrix at negative 
energy, both of them being real functions of momenta. A formula for the 
multipole polarizabilities of a two-body bound complex formed by a central 
interaction potential has been derived and studied. To test, the developed 
$T$-matrix formalism has been applied to the calculation of the dipole, 
quadrupole and octupole polarizabilities of the hydrogen atom.}\\[.2in]
{\footnotesize \it Keywords}: {\footnotesize Electric multipole polarizabilities;
few-body quantum systems; hydrogen atom} \\[.2in]
{\footnotesize PACS Nos.: 21.10.Ky; 21.45.+v; 32.10.Dk}
\end{abstract}

\vspace*{.1in}
\noindent {\bf 1. Introduction} \\ [.1in]
Study of the few-body quantum systems persists a perspective line of the 
advancement of the modern physics. In atomic physics, the Efimov effect 
for the three-body system, the existence of which was predicted in 1970 
(Ref. 1), has been experimentally verified using ultracold atoms and tuning 
the atom-atom scattering length by the magnetic field near a Feshbach resonance
$^2$ (see, for example, the reviews$^{3-5}$). In nuclear physics, extensive 
investigations of few-body nuclei advantageously undertaken over the past fifty 
years make it possible to gain new important information about the nuclear force. 

The investigation of the behaviour of the few-body nuclei in the external 
electromagnetic field permits to obtain additional data on their properties, 
specifically, on the electric polarizabilities and magnetic susceptibilities 
as important fundamental quantities of complex systems. In spite of the 
fact that the study of the electric polarizabilities of the few-body nuclei 
has attracted considerable interest from both the experimental$^{6-11}$ and 
theoretical$^{11-25}$ points of view and definite progress in the field has 
been reached, a great deal needs to be done in this area.
 
Up to now, there exists a discrepancy between the direct experimental result
for the electric dipole polarizability of the nucleus $^3$He (determined by 
measuring deviations from the Rutherford scattering law of the low-energy 
elastic $^3$He scattering by the Coulomb field of the heavy nucleus $^{208}$Pb 
(Ref. 8)) and the result deduced from the data for the total $^3$He 
photoabsorption cross section using the sum rule $\sigma_{-2}$ (Ref. 9). 

The conventional formula for the electric dipole polarizability of the 
$N$-particle bound complex has the form
\begin{equation}
\alpha_{E1} = 2 \sum_{\nu \neq 0}^{} \frac{\mid \langle \Psi_{\nu} \mid  
{\bf D}_1 \cdot \hat{{\mbox{\boldmath$\rho$}}}_0
\mid \Psi_0 \rangle \mid ^2}{E_{\nu}+B_0}\;,
\end{equation}
where ${\bf D}_1 = e_1 {\bf r}_1$ is the dipole moment operator, $E_{\nu}$ 
and $\Psi_{\nu}$ are the energy and the wave function of the $\nu$-excited state, 
${\bf r}_1$ and ${\mbox{\boldmath $\rho$}}_0$ are the radius vectors of the 
particle 1 relative to the center of mass of the bound complex and  
the center of mass of the complex relative to the charged particle $0$ 
creating the electric field, and the summation is taken over all possible 
excited discrete bound and continuum states. 

Although the formula (1) using the spectral expansion can be practically 
applied in the case of the two-body complex ($N=2$), the direct determination 
of $\alpha_{E1}$ for three and more body complexes ($N\geq 3$) by this 
relation with the performed beforehand calculation of all discrete and continuum 
states is not feasible. The known method of Dalgarno and Lewis$^{26}$ permits 
to bypass the difficulty related with taking account of the intermediate
continuum states in all possible channels in Eq.(1) introducing an additional 
function $\mid F \rangle$, 
\begin{displaymath}
\alpha_{E1} = 2 \langle \Psi_0 \mid {\bf D}_1 \cdot \hat{{\mbox{\boldmath$\rho$}}}_0
\mid F \rangle \;\;, \qquad  \mid F \rangle =
\sum_{\nu \neq 0}^{} \frac{\mid \Psi_{\nu} \rangle \langle \Psi_{\nu} \mid  
{\bf D}_1 \cdot \hat{{\mbox{\boldmath$\rho$}}}_0 \mid \Psi_0 \rangle}{E_{\nu}+B_0}\;,
\end{displaymath}
where $\mid F \rangle$ being the first-order correction to the unperturbated state 
satisfies an inhomogeneous differential equation. 
Dalgarno-Lewis method has been effectively applied in Refs. 12 and 23 to calculate 
the electric polarizabilities of hydrogen and helium isotopes. Though 
the result for $\alpha_{E1}(^3\mbox{He})$ obtained therewith supports the value 
deduced using the sum rule $\sigma_{-2}$ (Ref. 9), the subject 
of the discrepancy of the experimental data for $^3$He under discussion 
may not be considered as conclusively established since indispensable 
consistent calculations of the electric dipole polarizabilities of the 
three-body nuclei on the basis of the rigorous mathematical Faddeev's$^{27}$ 
formalism have not be performed yet.

In this paper we formulate the direct $T$-matrix approach (DTMA) to 
determination of the electric polarizabilities of a bound system,
that relies on the solution of the integral equations for both the 
bound-state wave function and corresponding components of the 
partial transition matrix of the system. In Section 2, following 
the Watson-Feshbach method, we express the polarization potential 
describing interaction between a charge particle and a bound complex 
consisting of $N$ particles in terms of a "truncated" Green's operator 
of the system. In this way we obtain a general expression for the electric 
multipole polarizability of the system. In Section 3 the $T$-matrix
approach to determination of the electric polarizabilities is
formulated and simplifications of the general formula assuming  
conservation of the space parity and the total orbital moment 
of the system, are considered. Section 4 is devoted to the application 
of the elaborated formalism to the two-body bound systems with the central
interaction. It is shown that the electric $2^{\lambda}$-pole 
polarizability of the two-particle $S$-state bound complex 
contains information both on derivatives (of the order $\lambda$ 
and lower) of the wave function and on the partial component of the 
transition matrix that corresponds to the orbital state with $l=\lambda$. 
Section 5 contains the application of the T-matrix formalism to determination 
of the electric multipole polarizabilities of the hydrogen atom, 
discussion and conclusions. \\

\vspace*{.1in}
\noindent {\bf 2. Polarization potential} \\ [.1in]
The formula for the polarization potential describing the interaction
between a charged particle $0$ and a bound complex consisting of $N$ 
particles follows imediately when treating the $N+1$ body problem, 
namely, the low-energy scattering of the complex by the Coulomb field 
of the charged particle $0$ with the kinetic energy of the relative motion 
of the particle $0$ and the complex $\epsilon$ being well below the breakup 
threshold energy of the complex. Experimentally, to determine the 
polarizability of the complex in the direct way a heavy nucleus (the 
particle $0$) is used as a source of the intense electric field. 
For simplicity sake assume that the $N$-body complex contains only one 
charged particle $1$, the other particles ($2,3,\ldots N$) being neutral.

The Hamiltonian of whole system of $N+1$ particle has the form 
\begin{equation}
{\cal H}=h^{\circ}_0 + H + v^C_{01}\;,
\end{equation}
where $h^{\circ}_0$ is the operator of the kinetic energy of the relative
motion of the particle $0$ and the center of mass of the complex, $H$ is the 
total Hamiltonian of the $N$-body complex, $H=H^{\circ} + V$, $H^{\circ}$ is 
the kinetic energy operator of the relative motion of the particles inside 
the complex, $V$ is the total interaction potential of the complex, 
$V = \sum_{i<j}^{} v_{ij}$ , $v_{01}^C$ is the operator of the Coulomb 
interaction between the charge particles $0$ and $1$ (with the charges 
$e_0$ and $e_1$), $v_{ij}$ is the operator of the interaction between 
the particles of the complex $i$ and $j$.

Using the known Watson-Feshbach projection technique$^{2,29,30}$ with the
projection operators $P = \mid \Psi_0><\Psi_0\mid$ and  $Q = 1 - P$, where  
$\Psi_0$ is the wave function of the ground bound state of the complex of $N$          
particles ($1,2,\ldots N$) with the binding energy $B_0$ (normalized to one),
the operator of the effective interaction may be written as 
\begin{equation}
V_{eff}({\varepsilon}) = < \Psi_0 \mid {\cal R}({\cal E}) \mid \Psi_0 >\;,
\end{equation}
where the averaging is taken over the variables of the relative motion of the
particles of the complex. The operator  ${\cal R}({\cal E})$ satisfies the 
integral equation
\begin{equation}
{\cal R}({\cal E}) = v_{01}^C + v_{01}^C {\cal G}^Q_{int} ({\cal E}){\cal R}({\cal E})\;,
\end{equation}
in which the potential energy operator is the "external" Coulomb interaction
potential between the field source (the particle $0$) and the charged particle $1$
of the complex, $v_{01}^C$, and the propagator is a "truncated" Green's operator
containing the "internal" interaction potential of the complex $V$,
\begin{equation}
{\cal G}^Q_{int} ({\cal E}) = Q {\cal G}_{int}({\cal E}), \qquad {\cal G}_{int} ({\cal E}) = 
({\cal E} - {\cal H}^{\circ} - V)^{-1},
\end{equation}
where ${\cal H}^{\circ}$ is the free Hamiltonian of the complete system 
(of $N+1$ particles), ${\cal H}^{\circ} = h^{\circ}_0 + H^{\circ}$, ${\cal E}$ is 
the total energy of the system, ${\cal E} = \varepsilon - B_0$.
                           
Introducing the transition operator ${\cal T}({\cal E})$ which satisfies the integral
Lippmann-Schwinger equation                             
\begin{equation}
{\cal T}({\cal E}) = V + V {\cal G}^{\circ} ({\cal E}){\cal T}({\cal E}),
\end{equation}
the operator ${\cal G}_{int} ({\cal E})$ can be written in the form
\begin{equation}
{\cal G}_{int} ({\cal E}) = {\cal G}^{\circ}({\cal E}) + {\cal G}^{\circ}({\cal E})
 {\cal T}({\cal E}) {\cal G}^{\circ}({\cal E}),
\end{equation}
where the free propagator ${\cal G}^{\circ}$ is given by     
${\cal G}^{\circ}({\cal E}) = ({\cal E} - {\cal H}^{\circ})^{-1}$.
   
The polarization potential that corresponds to the second order of the perturbation
expansion (in powers of the Coulomb interaction $v_{01}^C$) of the operator 
${\cal R}({\cal E})$ in the expression for the effective potential (3) is given by
\begin{equation}
V_{pol}(\varepsilon) =  < \Psi_0 \mid v_{01}^C {\cal G}^Q_{int}({\cal E}) v_{01}^C 
\mid \Psi_0 >, 
\end{equation}
where
\begin{equation}
{\cal G}_{int}^Q ({\cal E}) = {\cal G}^{\circ}({\cal E}) + {\cal G}^{\circ}({\cal E})
 {\cal T}({\cal E}) {\cal G}^{\circ}({\cal E}) -P {\cal G}_{int}({\cal E}).
\end{equation}

According to the uncertainty principle, in the case of asymptotically large (in 
comparison with the complex size) distances between the particle $0$ and the complex,
$\rho_0$, the momentum variables of relative motion among particles inside the complex 
considerably exceed the momentum variable of relative motion of the particle $0$ 
and the center of mass of the complex. In such a case, in the expression for 
the "truncated" Green's function ${\cal G}^Q_{int}({\cal E})$ in Eq.(8)
we can neglect with the variable quantity that corresponds to the operator  
$\varepsilon - h^{\circ}_0$ in comparison to the variable quantity that corresponds 
to the operator $H^{\circ} + B_0$. Then the polarization potential takes the known 
local form   
\begin{equation}
V_{pol}(\rho_0) = - \frac{e_0^2}{2} \sum_{\lambda = 1}^{\infty} 
\frac{\alpha_{E\lambda}}{\rho_0^{2\lambda+2}}\;,
\end{equation}
where each of coefficients in the sum (10), $\alpha_{E\lambda}$, that characterizes  
the strength of the individual constituent of the polarization potential
with the asymptotic  ${\rho_0^{-2\lambda-2}}$ is the electric polarizability of the  $N$-particle
complex of the multipolarity $2^{\lambda}$       
\begin{equation}
\alpha_{E\lambda}=-2<\Psi_0\mid M_{\lambda} G^Q(-B_0) M_{\lambda}\mid \Psi_0>\,.
\end{equation}
Here $M_{\lambda}$ is the multipole moment operator of the charged constituent 
particle $1$,    
\begin{equation}
M_{\lambda}=e_1 r_1^{\lambda} P_{\lambda}(\hat{{\bf r}}_1 \cdot \hat{{\mbox{\boldmath$\rho$}}}_0),
\end{equation}
${\bf r}_1$ is the radius vector of the particle 1 relative to the center of mass of the bound
complex; the radius vector ${\mbox{\boldmath $\rho$}}_0$   describes the position 
of the center of mass of the bound complex relative to the charged particle $0$ 
creating the electric field, $P_{\lambda}(x)$ is Legendre polinomial (the unit 
vectors are marked with a hat, $\hat{\mbox{\boldmath $a$}}\equiv \mbox{\boldmath $a$}/a$); 
the quantity $G^Q(-B_0)$ is the "truncated" Green's operator of the complex $G^Q(E)=QG(E)$ 
at the bound state energy $E=-B_0$, $G(E)\equiv(E-H^{\circ}-V)^{-1}$ is the total Green's 
function of the complex.

Note that the formula (1) readily follows from the general formula (11) after 
applying the spectral expansion of the total Green's operator $G(-B_0)$ in the 
complete set of the eigenfunctions of the total $N$-particle Hamiltonian $H$.
\pagebreak
\\
\vspace*{.1in}

\noindent {\bf 3. $T$-matrix method of determination of the electric multipole\\ 
polarizabilities of a bound complex} \\ [.1in]
In this Section we formulate a method for the direct calculation of the 
polarizability of a few-particle bound system starting immediately
from the definition (11) without recourse to expansion in terms of
a set of excited discrete and continuum wave functions. 

The general formula (11) for the electric dipole polarizability
of a bound complex of $N$ particles ($1,2,\ldots N$) contains the 
ground-state wave function $\Psi_0$ and "truncated" propagator 
\begin{equation}
G^Q(E)=G(E)-PG(E)\;
\end{equation}
at the energy of the bound state $E=-B_0$. The wave function  $\Psi_0$         
satisfies the Schr\"{o}dinger equation 
\begin{displaymath}
(H_0+V+B_0)\Psi_0=0    
\end{displaymath}
or the equivalent homogeneous integral equation 
\begin{equation}
\Psi_0=G^{\circ}(-B_0)V\Psi_0.
\end{equation}

We shall restrict our consideration to the case that the complex is
in the ground bound state.  

The "truncated" propagator $G^Q(E)$, as distinct from the total one,     
 $G(E)$ does not contain the pole singularity at $E=-B_0$. In an explicit
form, we find the expression for $G^Q(E)$ writing the total Green
operator $G(E)$ in terms of the free propagator $G^{\circ}(E)=(E-H^{\circ})^{-1}$                
and the transition operator $T(E)$,  
\begin{equation}
G(E)=G^{\circ}(E)+G^{\circ}(E)T(E)G^{\circ}(E),                          
\end{equation}
where the operator $T(E)$ is determined by the Lippman-Schwinger integral equation
\begin{equation}
T(E)=V+VG^{\circ}(E)T(E)\;.
\end{equation}
In view of Eq.(15) and the equality
\begin{equation}
PG(E) = \frac{1}{E + B_0} P \;,
\end{equation}
the expression (13) takes the form 
\begin{equation}
G^Q(E) = G^{\circ}(E) + G^{\circ}(E) T(E) G^{\circ}(E)- \frac{1}{E + B_0} P\;.   
\end{equation}

Separating out from the total transition operator $T(E)$ the singular pole
part that corresponds to the ground bound state of the system at the
energy $E=-B_0$, we write 
\begin{equation}
T(E) = \frac{\mid\Gamma_0><\Gamma_0\mid}{E + B_0} + \tilde{T}(E)\;,
\end{equation}
where the vertex function $\mid\Gamma_0\rangle$ is expressed 
through the wave function of the ground bound state of the system, 
\begin{equation}
\mid\Gamma_0\rangle=\left[ G^{\circ}(-B_0) \right] ^{-1}\mid\Psi_0\rangle =
V\mid\Psi_0\rangle\;, 
\end{equation}
satisfying the homogeneous integral equation 
\begin{equation}
\mid \Gamma_0 \rangle = V G^{\circ}(-B_0) \mid \Gamma_0 \rangle\;, 
\end{equation}
and $\tilde{T}(E)$ denotes the smooth (non-singular) part of the transition
operator.

Taking into account the expression (19) we write the formula (18)
for the operator $G^Q(E)$in the form 
\begin{displaymath}
G^Q(E) = G^{\circ}(E) + G^{\circ}(E) \frac{\left[ G^{\circ}(-B_0) \right] ^{-1} \mid \Psi_0 \rangle 
\langle \Psi_0 \mid \left[ G^{\circ}(-B_0) \right] ^{-1}}{E + B_0} G^{\circ}(E) \\ [-2mm]
\end{displaymath}
\begin{equation}
- \frac{\mid \Psi_0 \rangle \langle \Psi_0 \mid}{E + B_0} + G^{\circ}(E) \tilde{T}(E) G^{\circ}(E)\;\;.
\end{equation}

Cancelling the pole terms in Eq.(22) with the use of the identity 
\begin{equation}
G^{\circ}(E) \left[ G^{\circ}(-B_0) \right] ^{-1} = \left[ G^{\circ}(-B_0) \right] ^{-1} G^{\circ}(E) 
= 1 - (E + B_0) G^{\circ}(E) 
\end{equation}
we may write the expression (22) in the form 
\begin{equation}
\begin{array}{rcl}
G^Q(E)& = &G^{\circ}(E)-G^{\circ}(E)\mid\Psi_0\rangle\langle\Psi_0\mid
-\mid\Psi_0\rangle\langle\Psi_0\mid G^{\circ}(E) \\[2mm]  \nonumber
& + &(E+B_0)G^{\circ}(E)\mid\Psi_0\rangle\langle\Psi_0\mid G^{\circ}(E)
+G^{\circ}(E)\tilde{T}(E) G^{\circ}(E)\;.  
\end{array}
\end{equation}

At the point of the negative energy of the bound state of the $N$-particle
complex,  $E=-B_0$, the operator $G^Q(-B_0)$, that is contained in the general 
formula for the electric dipole polarizability (11), becomes 
\begin{equation}
\begin{array}{rcl}
G^Q(-B_0) = G^{\circ}(-B_0)& - &G^{\circ}(-B_0) \mid \Psi_0 \rangle \langle \Psi_0 \mid 
- \mid \Psi_0 \rangle \langle \Psi_0 \mid G^{\circ}(-B_0) \\[2mm]  \nonumber
& + &G^{\circ}(-B_0)\tilde{T}(-B_0) G^{\circ}(-B_0)\;.  
\end{array}
\end{equation}

With the use of the expression (25) the formula for the polarizability (11) 
takes the form
\begin{equation}
\begin{array}{rcl}
\alpha_{E\lambda} & = & -\;\;2 \langle \Psi_0 \mid M_{\lambda}
G^{\circ}(-B_0) M_{\lambda} \mid \Psi_0 \rangle \\[3mm] \nonumber
\; &  & +\;\;2 \langle \Psi_0 \mid M_{\lambda} 
G^{\circ}(-B_0) \mid \Psi_0 \rangle \langle \Psi_0 \mid M_{\lambda} 
\mid \Psi_0 \rangle \\[3mm] 
\; &   & +\;\;2 \langle \Psi_0 \mid M_{\lambda} 
\mid \Psi_0 \rangle \langle \Psi_0 \mid G^{\circ}(-B_0) M_{\lambda} 
\mid \Psi_0 \rangle \\[3mm] \nonumber
\; &   & -\;\;2 \langle \Psi_0 \mid M_{\lambda} 
G^{\circ}(-B_0) \tilde{T}(-B_0) G^{\circ}(-B_0) 
M_{\lambda} \mid \Psi_0 \rangle\;.  \nonumber
\end{array}
\end{equation}

A noticeable simplification of the general formula for the electric 
multipole polarizability (26) takes place, if the interaction 
potential $V$ is invariant relative to the space reflaction
(for example, for the systems with the Coulomb or nuclear 
interactions). In this case the wave function of the bound
complex is characterized by a definite parity. The conservation
of the parity leads to nullification of the matrix elements 
\begin{equation}
\langle\Psi_0\mid M_{\lambda}\mid\Psi_0\rangle\;,\quad 
\langle\Psi_0\mid G^{\circ}(-B_0) M_{\lambda}\mid\Psi_0\rangle \mbox{ and }
\langle\Psi_0\mid M_{\lambda}G^{\circ}(-B_0)\mid\Psi_0\rangle\;, 
\end{equation}
which are present in the second and third summands of the 
formula (26), at odd values of $\lambda$.
 
The formula (26) is also simplified in the case of the invariance 
of the interaction relative to rotations that leads to conservation 
of the angular momentum. In such a situation the matrix elements (27) 
are proportional to the Clebsh-Gordan coefficient 
$(L\lambda M\mu\mid LM^{\prime})$, satisfying the triangle 
condition $\Delta(L\lambda L)$. Specifically, they vanish 
after integrating in angular variables for all $\lambda\neq 0$, 
if the total orbital moment of the complex is equal to zero, $L=0$.

In the case when the ground bound state of the complex
is characterized with the total orbital momentum $L=0$,
that is real for the symplest atomic and few-hadron systems, 
the formula for the electric multipole polarizability (26) 
is simplified to the expression 
\begin{equation}
\begin{array}{rcl}
\alpha_{E\lambda} & = & -\;\;2\langle\Psi_0\mid M_{\lambda} G^{\circ}(-B_0)
M_{\lambda}\mid\Psi_0\rangle \\[3mm] \nonumber
&   & -\;\;2\langle\Psi_0\mid M_{\lambda} G^{\circ}(-B_0) \tilde{T}(-B_0) G^{\circ}(-B_0)
M_{\lambda}\mid\Psi_0\rangle\;.  
\end{array}
\end{equation}

As is evident from (26) or (28), to determine the electric 
dipole polarizability of the bound complex, it is necessary 
to know not only its wave function, but the smooth part  of
the transition matrix at negative energy of the bound state
($E=-B_0$) as well. The corresponding transition matrix can be
determined by solving the Lippmann-Schwinger integral equation
for the two-particle system, the Faddeev integral
equations$^{27}$ for the three-particle system or the 
Faddeev-Yakubovsky equations$^{28}$ for more complex systems. \\
\vspace*{.1in}

\noindent {\bf 4. Electric multipole polarizabilities of the 
two-particle bound complex } \\ [.1in]
In the case of two-particle bound complexes, the derived 
formula for the electric multipole polarizability (28) is
simplified. Considered here is a stable bound complex consisting
of a charged particle $1$ and a neutral particle $2$. The interaction
potential between the particles is taken to be central, the complex
is in $S$-wave ground bound state. We denote quantities describing
two-particle system by small letters as distinct to notations by
capital letters of quantities for $N$-particle complexes used
in the foregoing Sections: $b_0=\hbar^2 \kappa_0^2/2\mu_{12}$ and $\psi_0$        
are the binding energy and the wave function of the two-particle
bound complex, $g^{\circ}(-b_0)=(-b_0 - h^{\circ})^{-1}$ is the free 
Green's operator, $t(-b_0)$ is the two-particle transition operator,                
$\mu_{12}=m_1 m_2/(m_1 + m_2)$ is the reduced mass of the particle $1$
and $2$. Further consideration we perform in the momentum space.
        
Starting from (28) and taking into consideration that the smooth part 
of the two-particle transition matrix $\tilde{t}(-b_0)$         
has form of the sum of the smooth part of its $S$-wave partial
component (with  $l=0$), $\tilde{t}_0(-b_0)$,  and the sum of all
higher partial orbital components (with $l>0$), $t^{(h)}(-b_0)$, 
which are non-singular at the energy of the bound state,
\begin{equation}
\tilde{t}(-b_0)=\tilde{t}_0(-b_0)+t^{(h)}(-b_0)\;,   
\end{equation}
we write the formula for the electric multipole polarizability
of the two-particle system in the form
\begin{equation}
\begin{array}{rcl}
\alpha_{E\lambda} & = & -\;\;2\langle\psi_0\mid M_{\lambda} g^{\circ}(-b_0)
M_{\lambda}\mid \psi_0\rangle \\[3mm] \nonumber
&   & -\;\;2\langle\psi_0\mid M_{\lambda} g^{\circ}(-b_0) \left[\tilde{t}_0(-b_0)+
t^{(h)}(-b_0) \right] g^{\circ}(-b_0) M_{\lambda}\mid \psi_0\rangle\;.  
\end{array}
\end{equation}

The wave function of the complex in the $S$-wave bound state satisfies 
the homogeneous integral equation
\begin{equation}
\mid \psi_0 \rangle = g^{\circ}(-b_0)v_0 \mid \psi_0 \rangle\;.
\end{equation}
and each of the partial components of two transition matrix
$t_l(\epsilon)$ satisfies the inhomogeneous integral Lippmann-
Schwinger equations
\begin{equation}
t_l(\epsilon)=v_l+v_l g^{\circ}(\epsilon)t_l(\epsilon)\;.
\end{equation}
where $v_l$ is the partial component of the interaction
potential,  $\epsilon$ is the energy of the relative motion of the
particles.

The solution of the equation for the partial transition matrix (32)
that corresponds to the orbital moment of the ground state
($(l=0)$) has the form of the sum of the pole and smooth 
operators
\begin{equation}
t_0(\epsilon) = \frac{\mid\gamma_0><\gamma_0\mid}{\epsilon+b_0}+\tilde{t}_0(\epsilon)\;,
\end{equation}
where the vertex function $\mid\gamma_0\rangle \equiv v_0 \mid \psi_0\rangle
=(-b_0 - h^{\circ}) \mid \psi_0 \rangle$ satisfies the homogeneous integral equations 
that follows from the equation (31),
\begin{equation}
\mid \gamma_0\rangle = v_0 g^{\circ}(-b_0) \mid \gamma_0\rangle\;.
\end{equation}

Using the operator  $\omega_0(\epsilon)\equiv(\epsilon+b_0)t_0(\epsilon)$, which 
satisfies the inhomogeneous equation
\begin{equation}
\omega_0(\epsilon)=(\epsilon + b_0)v_0 + v_0 g^{\circ}(\epsilon)\omega_0(\epsilon)\;.
\end{equation}
we write the smooth part of the transition operator, $\tilde{t}_0(\epsilon)$ in the form 
\begin{equation}
\tilde{t}_0(\epsilon)=\frac{\omega_0(\epsilon)-\mid \gamma_0\rangle\langle\gamma_0\mid }
{\epsilon + b_0}\;.
\end{equation}   
At the point $\epsilon=-b_0$, the inhomogeneous equation (35)becomes homogeneous one,
\begin{equation}
\omega_0(-b_0) = v_0 g^{\circ}(-b_0) \omega_0(-b_0)\;.
\end{equation}
Since the kernels of the equations for the function $\mid \gamma_0\rangle$ (34) 
and the operator $\omega_0(-b_0)$ (35) coincide, the solution of the operator 
equation (37) may be written as          
\begin{equation}
\omega_0(-b_0) = \mid \gamma_0\rangle\langle\gamma_0\mid \;.
\end{equation}

According to (36), the operator $\tilde{t}_0(-b_0)$, which is contained 
in the formula for polarizability (30), is related to the operator 
$\omega_0(\epsilon)$ by the expression
\begin{equation}
\tilde{t}_0(-b_0) = \left.\frac{\partial \omega_0(\epsilon)}{\partial \epsilon} 
                      \right|_{\epsilon=-b_0}\equiv \omega_0^{\prime}(-b_0)\;.
\end{equation}
The operator $\omega_0^{\prime}(\epsilon)$ can be deduced by performing 
the differentiation of the equation (35) with respect to $\epsilon$ and 
the inverted transition from $v$ to $t$ and $\omega_0$,
\begin{equation}
\omega_0^{\prime}(\epsilon)=\frac{\omega_0(\epsilon) + 
\omega_0(\epsilon) {g^{\circ}}^{\prime}(\epsilon)
\omega_0(\epsilon)}{\epsilon + b_0}\;,\quad {g^{\circ}}^{\prime}(\epsilon)=
-(\epsilon-h^{\circ})^{-2}\;.
\end{equation}   
Evaluating on the right-hand side of (40) the inderterminacy of 
the type $\frac{0}{0}$ at the point  $\epsilon=-b_0$, which appears in view 
of the relation (38) and the normalization condition
$\langle\gamma_0\mid {g^{\circ}}^{\prime}(-b_0)] \mid \gamma_0\rangle = 
-\langle \psi_0 \mid \psi_0 \rangle = -1$, we obtain the expression for 
the operator $\tilde{t}_0(-b_0)$,                                       
\begin{equation}
\tilde{t}_0(-b_0) = \omega_0^{\prime}(-b_0) =  \mid \gamma_0\rangle \left(-\frac{R_1}{b_0} \right)
\langle\gamma_0\mid\;.
\end{equation}
where
\begin{equation}
R_1 = -\frac{1}{2} b_0 \langle\gamma_0\mid {g^{\circ}}^{\prime\prime}(-b_0) \mid \gamma_0\rangle\;,
\quad {g^{\circ}}^{\prime\prime}(\epsilon)]=2(\epsilon-h^{\circ})^{-3}\;.
\end{equation}
In the case of the central interaction, the part of the transition 
matrix $t^{(h)}(-b_0)$ in (30) may be written in the form of expansion 
in the set of the spherical functions of the angular variable momenta 
(with $l \geq 1$), 
\begin{equation}
<{\bf k}|t^{(h)}(-b_0)|{\bf k^{\prime}}>=4\pi\sum_{l=1}^{\infty} \sum_{m=-l}^{l}
        t_l(k,k^{\prime}; -b_0) Y_{lm}(\hat{\bf k})Y^{*}_{lm}(\hat{\bf k}^{\prime})\;.
\end{equation}

Substituting the expressions (41) and (43) for $\tilde{t}_0(-b_0)$          
and  $t^{h}(-b_0)$ into the formula for the electric polarizability (30) 
and taking account of the action of the multipole moment operator 
$M_{\lambda}$ (11) on the function $\psi_0$ in the momentum space 
(with  ${\bf r}=i\hbar \nabla_{\bf k}$),   
\begin{equation}
\langle {\bf k} \mid M_{\lambda} \mid \psi_0 \rangle = 
i^{\lambda} e_1 \left( - \frac{m_1}{m_{12}} \right)^{\lambda} \varphi_{\lambda}(k) 
P_{\lambda}(\hat{{\bf k}} \cdot \hat{{\mbox{\boldmath $\rho$}}}_0) \;,
\end{equation}
where
\begin{equation}
\varphi_{\lambda}(k)= (-1)^{\lambda} k^{\lambda}\left[ \left( \frac{1}{k} \frac{d}{dk} \right)^{\lambda} 
\psi_0(k) \right], \; \qquad m_{12}=m_1 + m_2 \;,
\end{equation}
we perform integration with respect to the angular variables. 

Notice that the contribution of the smooth part of the $S$-wave components 
of the transition matrix, $\tilde{t}_0(-b_0)$, which is of the 
separable form (41), proves to be equal to zero as a result of the 
conservation of the space parity and the orbital moment of the relative motion 
(similar to zero contributions in the general case from other 
factorable terms of the "truncated" Green's operator (25), see 
comments below the formula (26)). Non vanishing contribution in (30) 
makes only the partial component with $l=\lambda$ from Eq.(43), 
$\; t_{\lambda}(k,k^{\prime}; -b_0)\;$ that satisfies the Lippmann-Schwinger 
integral equation (32),    
\begin{equation}
t_{\lambda}(k,k^{\prime};-b_0) = v_{\lambda}(k,k^{\prime})-
\int_{0}^{\infty} \frac{dk^{\prime\prime} k^{{\prime\prime}2}}{2\pi^2}
v_{\lambda}(k,k^{\prime\prime}) \frac{1}{\frac{k^{{\prime\prime}2}}{2\mu_{12}}+b_0}
t_{\lambda}(k^{\prime\prime},k^{\prime};-b_0)\;\;, \\[3mm]  
\end{equation}
where  $v_{\lambda}(k,k^{\prime})$ is partial component of the interaction potential. 

The final result for the electric multipole polarizability of the two-particle 
complex with a central interaction between particles following from Eq. (30)
is written in the form 
\pagebreak
\begin{displaymath}
\alpha_{E\lambda} = \frac{2}{(2\lambda+1)\pi^2} \frac{m_1{e_1}^2}{\hbar^2}
\left( \frac{m_2}{m_{12}} \right) ^{2\lambda+1} 
\left\{ \int_{0}^{\infty} dk k^2 \frac{\mid \varphi_{\lambda}(k)
\mid ^2}{k^2+\kappa_0^2} \right. \\ [-2mm] \nonumber
\end{displaymath} 
\begin{equation}
  - \frac{1}{\pi^2} \frac{m_1 m_2}{\hbar^2 m_{12}} \int_{0}^{\infty}dk k^2 
\int_{0}^{\infty} dk^{\prime}k^{{\prime}2} 
\left. \frac{\varphi^{*}_{\lambda}(k) t_{\lambda}(k,k^{\prime};-b_0)
\varphi_{\lambda}(k^{\prime})}
{(k^2+\kappa_0^2)(k^{{\prime}2}+\kappa_0^2)} \right\}\;. 
\end{equation}

According to (45), the functions $\varphi_{\lambda}(k)$ in (47) are expressed in 
terms of the derivatives of the wave function of the ground bound state
$\psi_0(k)$ with respect to the relative momentum variable $k$. In the 
cases of the dipole ($\lambda=1$), quadrupole ($\lambda=2$) and octupole ($\lambda=3$)
polarizabilities they are of the form   
\begin{equation}
\varphi_{\lambda}(k) = \left\{ \begin{array}{lr}
 - \psi_0^{\prime}(k) & \mbox{for $\lambda=1$}\;, \\ 
 + \psi_0^{\prime\prime}(k) - \frac{1}{k} \psi_0^{\prime}(k) & \mbox{for $\lambda=2$}\;, \\
 - \psi_0^{\prime\prime\prime}(k) + \frac{3}{k} \psi_0^{\prime\prime}(k) 
 - \frac{3}{k^2} \psi_0^{\prime}(k) & \mbox{for $\lambda=3$}\;.
                                \end{array}
                      \right.                                
\end{equation}

The formula (47) derived in the case of the central interaction 
between the constituents of the two-body complex demonstrates that 
the electric $2^\lambda$-pole polarizability of the complex 
contains information not only on the derivatives (of the order 
$\lambda$ and lower) of its wave function, but on the partial 
component of the transition matrix, $t_{\lambda}(k,k^{\prime};-b_0)$, 
as well.
                  
It is advantageous to write the formula (47) in the form more 
convenient for the practical use. In order to do this, instead of 
the two-particle $t$-matrix (that is the function of two momentum 
variables $k$ and $k^\prime$), we introduce in (47) the function of one 
variable                              
\begin{equation}
\phi_\lambda (k) = - \int_{0}^{\infty} \frac{dk^{\prime} k^{{\prime}2}}{2\pi^2}
t_{\lambda}(k^,k^{\prime};-b_0)\frac{1}{\frac{k^{{\prime}2}}{2\mu_{12}}+b_0} \varphi_{\lambda}(k^{\prime}) 
\end{equation}
that satisfies the inhomogeneous integral equation with the kernel 
of the Lippmann-Schwinger equation (46)
\begin{equation}
\phi_{\lambda}(k) = f_{\lambda}(k) - \int_{0}^{\infty} \frac{dk^{\prime} k^{{\prime}2}}{2\pi^2}
v_{\lambda}(k,k^{\prime})\frac{1}{\frac{k^{{\prime}2}}{2\mu_{12}}+b_0} \phi_{\lambda}(k^{\prime}) 
\end{equation}
in which the free term is determined by the formula
\begin{equation}
f_{\lambda}(k) = - \int_{0}^{\infty} \frac{dk^{\prime} k^{{\prime}2}}{2\pi^2}
v_{\lambda}(k, k^{\prime}) \frac{1}{\frac{k^{{\prime}2}}{2\mu_{12}}+b_0} \varphi_\lambda(k^{\prime}). 
\end{equation}
In such a case, the formula (47) takes the form
\begin{equation}
\alpha_{E\lambda} = \frac{2}{(2\lambda+1)\pi^2} \frac{m_1{e_1}^2}{\hbar^2}
\left(\frac{m_2}{m_{12}}\right)^{2\lambda+1}
\int_{0}^{\infty} dk k^2 \frac{\varphi_{\lambda}^{*}(k) \left[ \varphi_{\lambda}(k) 
- \phi_{\lambda}(k) \right]}{k^2 + \kappa_0^2} \;.
\end{equation}
\\
\vspace*{.1in}

\noindent {\bf 5. Application of the DTMA, discussion and conclusions} \\ [.1in]
Advantages of the $t$-matrix approach are manifested in calculation of 
polarizabilities of a quantum system, even if it is two-particle (the deuteron 
nucleus or the hydrogen atom). In the special case that the interaction 
between particles is purely $S$-wave (for example, the separable interaction 
potential$^{31}$), in the formula (47) only the first term persists. 
The corresponding reduced formula for the dipole polarizability of the 
two-particle bound complex has been earlier derived in the framework 
of the three-body formalism of the effective interaction of a charged particle 
and a complex$^{13,14,32}$. Applications of the $t$-matrix approach to 
calculation of the electric dipole polarizabilities of the deuteron, the triton 
and the lambda hypertriton as two cluster systems with purely $S$-wave interaction 
have been carried out in the preceding our papers$^{24,25}$.  

For the deuteron with noncentral interaction between the proton and 
the neutron (in the state with the total angular momentum 1) the 
polarization in the electric field is anisotropic. Using the 
separable tensor potential$^{33}$ the longitudinal and transverse 
(relatively to the direction of the electric field) components of the 
dipole polarizability of the deuteron have been calculated in the 
work$^{15}$. The results of the further calculations of the components 
of the deuteron electric dipole polarizability obtained in the 
framework of the chiral effective field theory$^{18}$ are in agreement with 
the results of Ref. 15.

In the case of the hydrogen atom H (assuming that the proton 
mass is infinitely great when compared to the electron mass), the 
exact values of the electric multipole polarizabilities 
$\alpha_{E\lambda}(\mbox{H})$ are known$^{26}$. This is why it is 
possible to test directly the validity of the general formula for 
the polarizability of the two-particle system (47) --- both of the term 
with free propagation in intermediate state and of the terms with the 
multiple scattering in the higher orbital intermediate states 
(with $l=1,2,3,\ldots$ for the dipole, quadrupole, octupole and higher 
multipole polarizabilities, respectively).

Inserting the analytical expression for the partial components of 
the Coulomb transition matrix, obtained from the representation for
the three-dimensional $t$-matrix $<{\bf k}|t^{C}(\epsilon)|{\bf k^{\prime}}>$ 
derived in Ref. 34 with the use of the O(4) rotation symmetry in four-dimensional 
Fock space$^{35}$, into the formula (47) and separating out the Born term 
from the Coulomb $t$-matrix,  
\begin{equation}
t_{\lambda}^C(k,k^{\prime};-b_0) = v_{\lambda}^C(k,k^{\prime})+
\Delta t_{\lambda}^C(k,k^{\prime};-b_0)\;\;,   
\end{equation}
we write $\alpha_{E\lambda}(\mbox{H})$ as a sum of three terms, 
\begin{equation}
\alpha_{E\lambda}(\mbox{H})=\alpha_{\lambda}^0 + \alpha_{\lambda}^B 
+ \alpha_{\lambda}^{MS}\;,
\end{equation}
where  $\alpha_{\lambda}^0$ is the first term in Eq.(47) that describes free 
virtual propagation, $\alpha_{\lambda}^B$ is a part of the second term that 
contains only the Born term of the partial Coulomb $t$-matrix, $v_{\lambda}^C$, 
describing the single scattering in the intermediate state, and $\alpha_{\lambda}^{MS}$ 
takes into account the multiple scattering contributions (of the order 2 and more).

We show in the Table 1 the values of of the components  
$\alpha_{\lambda}^0$, $\alpha_{\lambda}^B$, $\alpha_{\lambda}^{MS}$ and their 
sum (54) (in a. u.) that determines the electric dipole 
($\lambda=1$), quadrupole ($\lambda=2$), and octupole ($\lambda=3$) 
polarizabilities of the hydrogen atom, $\alpha_{E\lambda}(\mbox{H})$, obtained 
applying the direct $t$-matrix approach, together with the exact values of 
the polarizabilities derived by Dalgarno and Lewis$^{26}$. Here, values of the 
quantities $\alpha_{\lambda}^0$ and $\alpha_{\lambda}^B$ are derived from (47) 
analytically, and the values of $\alpha_{\lambda}^{MS}$ are obtained numerically 
calculating the integral with $\Delta t_{\lambda}^C(k,k^{\prime};-b_0)$ that describes 
the multiple scattering in intermediate states.
\pagebreak
\\
\vspace*{.1in} 

{\footnotesize \tablename\hspace{2mm}1.\hspace{1mm} The components 
$\alpha_{\lambda}^0$, $\alpha_{\lambda}^B$ and $\alpha_{\lambda}^{MS}$ 
determining the electric dipole ($\lambda=1$), quadrupole ($\lambda=2$) and octupole 
($\lambda=3$) polarizabilities of the hydrogen atom, $\alpha_{E\lambda}(\mbox{H})$, 
calculated with the use of DTMA (Eqs.(47), (53) and (54)) together with the exact 
values of the polarizabilities (taken from Ref. 26) (in a.u.)
\begin{center} \begin{tabular}{|c|c|c|c|c|c|} \hline
\multicolumn{1}{|c|}{}& \multicolumn{1}{c|}{}& \multicolumn{1}{c|}{}&
\multicolumn{1}{c|}{}&
\multicolumn{1}{c|}{}&
\multicolumn{1}{c|}{Exact} \\
\multicolumn{1}{|c|}{$\lambda$}&
\multicolumn{1}{c|}{$\alpha_{\lambda}^0$}&
\multicolumn{1}{c|}{$\alpha_{\lambda}^B$}&
\multicolumn{1}{c|}{$\alpha_{\lambda}^{MS}$}&
\multicolumn{1}{c|}{$\alpha_{E\lambda}(\mbox{H})$}&
\multicolumn{1}{c|}{value} \\
\multicolumn{1}{|c|}{}&
\multicolumn{1}{c|}{}&
\multicolumn{1}{c|}{}&
\multicolumn{1}{c|}{}&
\multicolumn{1}{c|}{}&
\multicolumn{1}{c|}{(Ref.26)} \\ \hline
\multicolumn{1}{|c|}{}&
\multicolumn{1}{c|}{}&
\multicolumn{1}{c|}{}&
\multicolumn{1}{c|}{}&
\multicolumn{1}{c|}{}&
\multicolumn{1}{c|}{} \\ 
$1$&$\frac{7}{3}$&$\frac{10}{9}$&$1.055555\ldots$&$4.499999\ldots$&$\frac{9}{2}$ \\
\multicolumn{1}{|c|}{}&
\multicolumn{1}{c|}{}&
\multicolumn{1}{c|}{}&
\multicolumn{1}{c|}{}&
\multicolumn{1}{c|}{}&
\multicolumn{1}{c|}{} \\ 
$2$&$\frac{81}{8}$&$\frac{105}{32}$&$1.593750$&$15.000000$&$15$\\
\multicolumn{1}{|c|}{}&
\multicolumn{1}{c|}{}&
\multicolumn{1}{c|}{}&
\multicolumn{1}{c|}{}&
\multicolumn{1}{c|}{}&
\multicolumn{1}{c|}{} \\ 
$3$&$99$&$\frac{243}{10}$&$7.950000$&$131.250000$&$\frac{525}{4}$ \\ 
\multicolumn{1}{|c|}{}&
\multicolumn{1}{c|}{}&
\multicolumn{1}{c|}{}&
\multicolumn{1}{c|}{}&
\multicolumn{1}{c|}{}&
\multicolumn{1}{c|}{} \\  \hline
\end{tabular}
\end{center}}
\bigskip

Data given in Table 1 indicate that the dipole polarizability of the
hydrogen atom consists of two nearly equal parts --- the first term 
with the free propagator, $\alpha_1^0=\frac{7}{3}$, and the sum 
of two others with the $P$-wave component of the Coulomb transition matrix,          
$\alpha_1^B + \alpha_1^{MS} = \frac{13}{6}$. For polarizabilities of 
higher polarity the contribution of the term with the free propagator 
$\alpha_{\lambda}^0$ still further increases reaching $67,5\%$ 
in the case $\lambda=2$ and $75,4\%$ 
in the case $\lambda=3$. The contribution of the Born term $\alpha_{\lambda}^B$      
to the polarizability $\alpha_{E\lambda}(\mbox{H})$ diminishes slowly with $\lambda$ 
being equal to $24.7\%$
for $\lambda=1$, $21.9\%$ 
for $\lambda=2$ and $18.5\%$
for $\lambda=3$. The contribution of the term describing multiple 
virtual scattering, $\alpha_{\lambda}^{MS}$, lowers with increasing 
$\lambda$ accounting for $23.5\%$ 
when $\lambda=1$, $10.6\%$
when  $\lambda=2$ and $6.1\%$ 
when $\lambda=3$. It is worthy of note that contribution of the sum of the 
analytically tractable terms, $\alpha_{\lambda}^0 + \alpha_{\lambda}^B$,
to the polarizabilities of the hydrogen atom $\alpha_{E\lambda}(\mbox{H})$ 
is prevailing, it accounts for $76.5\%$ 
(if $\lambda=1$), $89.4\%$ 
(if $\lambda=2$) and $93.9\%$ 
(if $\lambda=3$) of the total amount.
                                                      
The elaborated approach to determination of the electric multipole 
polarizabilities that relies on the two-particle $t$-matrix is 
found to be rather effective compared to the conventional method 
that uses the spectral expansion of the Green's function of the system. 
As evidenced by the results of the spectral-expansion calculations 
by Castillejo at al$^{36}$, $65.8\%$ 
of the magnitude of the electric dipole polarizability of the hydrogen 
atom comes from the $P$-wave excited bound states. Inclusion of the all 
excited bound states account for $81.4\%$
. 
The rest $18.6\%$ 
is provided taking into consideration the continuum states.  

To summarize, as one of advantages of the approach developed here is 
that there is no need to calculate the individual contributions from all 
discrete and continuum) excited states. Instead, the determination of 
one or several (depending on the interaction mode) partial components of 
the transition matrix is now required. It is also essential that the 
transition matrix in the main formula (28) depends on the negative energy 
of the bound state being therefore a real function of momenta. The developed 
method can be immediately extended to more complicated interactions between 
constituents, specifically, to the case of tensor interactions that 
gives rise anisotropic polarization properties of the system. 
Finally, it is quite important that the proposed method is suitable for 
more complicated three- and $N$-body systems described by the
Faddeev and Faddeev-Yakubovsky integral equations the kernels of which 
are expressed through components of the transition matrices for subsystems. 
In the first place, we plan to apply the T-matrix approach to study the 
deformation properties of the $^3$H and $^3$He nuclei in the electric field. \\

\vspace*{.1in}
\noindent {\footnotesize {\bf References}
\vspace*{.1in}
\begin{itemize}
\setlength{\baselineskip}{.1in}
\item[{\tt 1.}]V. Efimov, {\it Phys. Lett.} {\bf B33}, 563 (1970); Sov. J. Nucl. Phys. 
           {\bf 12}, 589 (1971) [{\it Yad. Fiz.} {\bf 12}, 1080 (1970)]. 
\item[{\tt 2.}]H. Feshbach, {\it Ann. Phys. (N.Y.)} {\bf 5}, 357 (1958);
           {\bf 19}, 287 (1962). 
\item[{\tt 3.}]E. Braaten and H.-W. Hammer, {\it Phys. Rept.} {\bf 428}, 
           259 (2006). 
\item[{\tt 4.}]T. Kraemer at el., {\it Nature (London)} {\bf 440}, 315 (2006).  
\item[{\tt 5.}]C. Chin, P. Grimm P. Julienne and E. Tiesinga, {\it Rev. Mod. Phys.} 
           {\bf 82}, 1225 (2010). 
\item[{\tt 6.}]N. L. Rodning, L. D.Knutson, W. G. Lynch, and M.
           B. Tsang, {\it Phys. Rev. Lett.} {\bf 49}, 909 (1982).
\item[{\tt 7.}]J. L. Friar, S. Fallieros, E. L. Tomusiak, D. Skopik and
           E. G. Fuller, {\it Phys. Rev.} {\bf C27}, 1364 (1983).           
\item[{\tt 8.}]F. Goeckner, L. O. Lamm and L. D. Knutson,
           {\it Phys. Rev.} {\bf C43}, 66 (1991).
\item[{\tt 9.}]G. A. Rinker, {\it Phys. Rev.} {\bf A14}, 18 (1976).
\item[{\tt 10.}]J. L. Friar, {\it Phys. Rev.} {\bf C16}, 1540 (1977).
\item[{\tt 11.}]K. Pachucki and A. M. Moro, {\it Phys. Rev.} {\bf A75}, 
           032521 (2007).
\item[{\tt 12.}]J. L. Friar and S. Fallieros, {\it Phys. Rev.} {\bf C29},
           232 (1984).
\item[{\tt 13.}]V. F. Kharchenko, S. A. Shadchin and S. A. Permyakov, {\it
           Phys. Lett.} {\bf B199}, 1 (1987).
\item[{\tt 14.}]V. F. Kharchenko and S. A. Shadchin, {\it Ukrainian J. Phys.}
           {\bf 42}, 912 (1997).
\item[{\tt 15.}]A. V. Kharchenko, {\it Nucl. Phys.} {\bf A617}, 34 (1997).
\item[{\tt 16.}]J. L. Friar and G. L. Payne, {\it Phys. Rev.} {\bf C55}, 2764
           (1997).
\item[{\tt 17.}]V. D. Efros, W.Leidemann and G. Orlandini, {\it Phys. Lett.}
           {\bf B408}, 1 (1997); nucl-th/9707007.           
\item[{\tt 18.}]J.-W. Chen, H. W. Grie{\ss}hammer, M. J. Savage and
           R. P. Springer, {\it Nucl. Phys.} {\bf A644}, 221 (1998);
           nucl-th/9806080.          
\item[{\tt 19.}]D. R. Phillips, G. Rupak and M. J. Savage, {\it Phys. Lett} 
           {\bf B473}, 209 (2000).
\item[{\tt 20.}]X. Ji and Y. Li, {\it Phys. Lett.} {\bf B591}, 76 (2004).           
\item[{\tt 21.}]W. Leidemann, in {\it Few-Body Problems in Physics '02}, edited by 
           R. Krivec, B. Golli, M. Rosina and S. Sirca (Springer-Verlag Wien, New-York, 
           2003), Vol. {\bf 14}, p. 313.
\item[{\tt 22.}]D. Gazit, N. Barnea, S. Bacca, W. Leidemann and G. Orlandini, 
           {\it Phys. Rev.} {\bf C74}, 061001 (2006).            
\item[{\tt 23.}]I. Stetcu, S. Quaglioni, J. L. Friar, A. C. Hayes and P. Navr$\acute{a}$til, 
           {\it Phys. Rev.} {\bf C79}, 064001 (2009).             
\item[{\tt 24.}]V. F. Kharchenko and A. V. Kharchenko, {\it Collected Physical
           Papers (Lviv)} {\bf 7}, 432 (2008); nucl-th/0811.2565.           
\item[{\tt 25.}]V. F. Kharchenko and A. V. Kharchenko, {\it Int. J. Mod. Phys.} {\bf E19}, 
           225 (2010); nucl-th/1003.5769. 
\item[{\tt 26.}]A. Dalgarno and J. T. Lewis, {\it Proc. Roy. Soc. (London)} {\bf A233}, 
           70 (1955).            
\item[{\tt 27.}]L. D. Faddeev, {\it Sov. Phys. JETP} {\bf 12}, 1014 (1961) 
           [{\it Zh. Eksp. Teor. Fiz.} {\bf 39}, 1459 (1960)].
\item[{\tt 28.}]O. A. Yakubovsky, Sov. J. Nucl. Phys. {\bf 5}, 937 (1967) [{\it Yad. Fiz.} 
           {\bf 5}, 1312 (1967)].
\item[{\tt 29.}]N. C. Francis and K. M. Watson, {\it Phys. Rev.} {\bf 92}, 291 (1953).
\item[{\tt 30.}]C. J. Joachain, {\it Quantum Collision Theory} (North-Holland --- American Elsevier, 
           Amsterdam --- New-York, 1975).           
\item[{\tt 31.}]Y. Yamaguchi, {\it Phys. Rev.} {\bf 95}, 1628 (1954).
\item[{\tt 32.}]V. F. Kharchenko and S. A. Shadchin S.A., {\it Three-body
           theory of the effective interaction between a particle and a
           two-particle bound system}, preprint ITP-93-24E (Institute for
           Theoretical Physics, Kyiv, 1993).
\item[{\tt 33.}]Y. Yamaguchi and Y. Yamaguchi, {\it Phys. Rev.} {\bf 95}, 1635 (1954).           
\item[{\tt 34.}]S. A. Shadchin and V. F. Kharchenko, {\it J. Phys.} 
           {\bf 16}, 1319 (1983).          
\item[{\tt 35.}]V. A. Fock, {\it Z. Phys.} {\bf 98}, 145 (1935).
\item[{\tt 36.}]L. Castillejo, I. C. Percival and M. J. Seaton, {\it Proc. Roy. Soc. 
           (London)} {\bf A254}, 259 (1960).            
           
\end{itemize}}

\end{document}